\documentstyle[prl,epsf,aps,floats]{revtex}  % use for PRL mode
\input psfig.sty
\def\ppbar{$p\overline{p}~$}             %ppbar
\def\et{$E_{\rm T}$}                          %ET
\def\met{\mbox{${\hbox{$E$\kern-0.6em\lower-.1ex\hbox{/}}}_T$}} %missing ET
\def\mex{\mbox{${\hbox{$E$\kern-0.6em\lower-.1ex\hbox{/}}}_x$}} %missing Ex
\def\mey{\mbox{${\hbox{$E$\kern-0.6em\lower-.1ex\hbox{/}}}_y$}} %missing Ey
\def\mexy{\mbox{${\hbox{$E$\kern-0.6em\lower-.1ex\hbox{/}}}_{x,y}$}} %missing Exy
 
\def\gevcc{GeV/$c^2~$}                   %GeV/c^2
\def\gevc{GeV/$c$}                       %GeV/c
\def\D0{D\O}                            %D0
\begin{document}
\lefthyphenmin=2
\righthyphenmin=3

%new 060600
%\setlength{\baselineskip}{24pt}
%\preprint{FERMILAB-Pub-97/109-E}
%
% ======> Title of the paper goes here <====================
%

\title{
Search for Excited and Exotic Electrons in the $e \gamma$ Decay Channel \\ 
in $\boldmath p\overline{p}$ Collisions at $\boldmath \sqrt{s} = 1.96$ TeV
}

\maketitle

\font\eightit=cmti8
\def\r#1{\ignorespaces $^{#1}$}
\hfilneg
\begin{sloppypar}
\noindent
D.~Acosta,\r {16} J.~Adelman,\r {12} T.~Affolder,\r 9 T.~Akimoto,\r {54}
M.G.~Albrow,\r {15} D.~Ambrose,\r {43} S.~Amerio,\r {42}  
D.~Amidei,\r {33} A.~Anastassov,\r {50} K.~Anikeev,\r {31}
% ...
\end{sloppypar}
\vskip .026in
\begin{center}
(CDF Collaboration)
\end{center}

\vskip .026in
\begin{center}
\r 1  {\eightit Institute of Physics, Academia Sinica, Taipei, Taiwan 11529, 
Republic of China} \\
\r 2  {\eightit Argonne National Laboratory, Argonne, Illinois 60439} \\
\r 3  {\eightit Institut de Fisica d'Altes Energies, Universitat Autonoma
de Barcelona, E-08193, Bellaterra (Barcelona), Spain} \\
\r 4  {\eightit Istituto Nazionale di Fisica Nucleare, University of Bologna,
I-40127 Bologna, Italy} \\
\r 5  {\eightit Brandeis University, Waltham, Massachusetts 02254} \\
\r 6  {\eightit University of California at Davis, Davis, California  95616} \\
\r 7  {\eightit University of California at Los Angeles, Los 
Angeles, California  90024} \\
\r 8  {\eightit University of California at San Diego, La Jolla, California  92093} \\ 
\r 9  {\eightit University of California at Santa Barbara, Santa Barbara, California 
93106} \\ 
\r {10} {\eightit Instituto de Fisica de Cantabria, CSIC-University of Cantabria, 
39005 Santander, Spain} \\
\r {11} {\eightit Carnegie Mellon University, Pittsburgh, PA  15213} \\
\r {12} {\eightit Enrico Fermi Institute, University of Chicago, Chicago, 
Illinois 60637} \\
\r {13}  {\eightit Joint Institute for Nuclear Research, RU-141980 Dubna, Russia}
\\
\r {14} {\eightit Duke University, Durham, North Carolina  27708} \\
\r {15} {\eightit Fermi National Accelerator Laboratory, Batavia, Illinois 
60510} \\
\r {16} {\eightit University of Florida, Gainesville, Florida  32611} \\
\r {17} {\eightit Laboratori Nazionali di Frascati, Istituto Nazionale di Fisica
               Nucleare, I-00044 Frascati, Italy} \\
\r {18} {\eightit University of Geneva, CH-1211 Geneva 4, Switzerland} \\
\r {19} {\eightit Glasgow University, Glasgow G12 8QQ, United Kingdom}\\
\r {20} {\eightit Harvard University, Cambridge, Massachusetts 02138} \\
\r {21} {\eightit The Helsinki Group: Helsinki Institute of Physics; and Division of
High Energy Physics, Department of Physical Sciences, University of Helsinki, FIN-00044, Helsinki, Finland}\\
\r {22} {\eightit Hiroshima University, Higashi-Hiroshima 724, Japan} \\
\r {23} {\eightit University of Illinois, Urbana, Illinois 61801} \\
\r {24} {\eightit The Johns Hopkins University, Baltimore, Maryland 21218} \\
\r {25} {\eightit Institut f\"{u}r Experimentelle Kernphysik, 
Universit\"{a}t Karlsruhe, 76128 Karlsruhe, Germany} \\
\r {26} {\eightit High Energy Accelerator Research Organization (KEK), Tsukuba, 
Ibaraki 305, Japan} \\
\r {27} {\eightit Center for High Energy Physics: Kyungpook National
University, Taegu 702-701; Seoul National University, Seoul 151-742; and
SungKyunKwan University, Suwon 440-746; Korea} \\
\r {28} {\eightit Ernest Orlando Lawrence Berkeley National Laboratory, 
Berkeley, California 94720} \\
\r {29} {\eightit University of Liverpool, Liverpool L69 7ZE, United Kingdom} \\
\r {30} {\eightit University College London, London WC1E 6BT, United Kingdom} \\
\r {31} {\eightit Massachusetts Institute of Technology, Cambridge,
Massachusetts  02139} \\   
\r {32} {\eightit Institute of Particle Physics: McGill University,
Montr\'{e}al, Canada H3A~2T8; and University of Toronto, Toronto, Canada
M5S~1A7} \\
\r {33} {\eightit University of Michigan, Ann Arbor, Michigan 48109} \\
\r {34} {\eightit Michigan State University, East Lansing, Michigan  48824} \\
\r {35} {\eightit Institution for Theoretical and Experimental Physics, ITEP,
Moscow 117259, Russia} \\
\r {36} {\eightit University of New Mexico, Albuquerque, New Mexico 87131} \\
\r {37} {\eightit Northwestern University, Evanston, Illinois  60208} \\
\r {38} {\eightit The Ohio State University, Columbus, Ohio  43210} \\  
\r {39} {\eightit Okayama University, Okayama 700-8530, Japan}\\  
\r {40} {\eightit Osaka City University, Osaka 588, Japan} \\
\r {41} {\eightit University of Oxford, Oxford OX1 3RH, United Kingdom} \\
\r {42} {\eightit University of Padova, Istituto Nazionale di Fisica 
          Nucleare, Sezione di Padova-Trento, I-35131 Padova, Italy} \\
\r {43} {\eightit University of Pennsylvania, Philadelphia, 
        Pennsylvania 19104} \\   
\r {44} {\eightit Istituto Nazionale di Fisica Nucleare, University and Scuola
               Normale Superiore of Pisa, I-56100 Pisa, Italy} \\
\r {45} {\eightit University of Pittsburgh, Pittsburgh, Pennsylvania 15260} \\
\r {46} {\eightit Purdue University, West Lafayette, Indiana 47907} \\
\r {47} {\eightit University of Rochester, Rochester, New York 14627} \\
\r {48} {\eightit The Rockefeller University, New York, New York 10021} \\
\r {49} {\eightit Istituto Nazionale di Fisica Nucleare, Sezione di Roma 1,
University di Roma ``La Sapienza," I-00185 Roma, Italy}\\
\r {50} {\eightit Rutgers University, Piscataway, New Jersey 08855} \\
\r {51} {\eightit Texas A\&M University, College Station, Texas 77843} \\
\r {52} {\eightit Texas Tech University, Lubbock, Texas 79409} \\
\r {53} {\eightit Istituto Nazionale di Fisica Nucleare, University of Trieste/\
Udine, Italy} \\
\r {54} {\eightit University of Tsukuba, Tsukuba, Ibaraki 305, Japan} \\
\r {55} {\eightit Tufts University, Medford, Massachusetts 02155} \\
\r {56} {\eightit Waseda University, Tokyo 169, Japan} \\
\r {57} {\eightit Wayne State University, Detroit, Michigan  48201} \\
\r {58} {\eightit University of Wisconsin, Madison, Wisconsin 53706} \\
\r {59} {\eightit Yale University, New Haven, Connecticut 06520} \\
\end{center}
 
%
% ==============> Text of the abstract goes here <=====================
% 
\begin{abstract}
 We present a search for excited and exotic electrons ($e^*$) decaying to
  an electron and a photon, both with high transverse momentum. We use 202 
pb$^{-1}$ of data collected in 
 {\mbox{$p\bar p$}}\ collisions at {\mbox{$\sqrt{s}$ =\ 1.96\ TeV }}
 with the CDF II detector.  
 No signal above standard model expectation is seen for associated $ee^*$
 production. 
 We discuss the $e^*$ sensitivity in the parameter space of the excited
  electron mass $M_{e^*}$ and
 the compositeness energy scale $\Lambda$. In the contact interaction
 model, we exclude 132~\gevcc\ 
 $< M_{e^*} < 879$~\gevcc\ for $\Lambda = M_{e^*}$
 at 95\% confidence level (C.L.). In the 
 gauge-mediated model, we exclude 126~\gevcc\ 
 $< M_{e^*} < 430$~\gevcc at 95\% C.L. 
 for the phenomenological coupling $f/\Lambda \approx 10^{-2}$ GeV$^{-1}$. 

\end{abstract}

\pacs{PACS numbers: 12.60.Rc, 13.85.Qk, 12.60.-i, 14.60.Hi}

%\newpage
\vfill\eject
\twocolumn

 The particle content of the standard model (SM) is given by three
 generations of quarks and leptons, each containing an $SU(2)$ 
 doublet. This fermion multiplicity motivates a description in terms
 of underlying substructure, in which all quarks and leptons consist of
 fewer elementary particles bound by a new strong 
 interaction~\cite{baurestar}. In this
 compositeness model, quark-antiquark annihilations 
 may result in the production of excited lepton states, such
 as the excited electron, $e^*$. 
 The SM  may  be embedded in larger gauge groups
 such as $SO(10)$ or $E(6)$, motivated by grand unified theories or
 string theory. These embeddings also 
 predict exotic fermions such
 as the $e^*$, produced  via
 their gauge interactions~\cite{baurestar}. 

 We search for associated $e e^*$ production followed by the radiative decay
 $e^* \rightarrow e \gamma$. This mode yields the distinctive $e e \gamma$ 
 final state, which is fully reconstructable with high efficiency and good 
 mass resolution, and has small backgrounds. The evidence for $e^*$
 production would be the observation of a resonance in the 
 $e \gamma$ invariant mass distribution. 
 The contact interaction (CI) Lagrangian~\cite{baurestar}
   describing the reaction $q \bar{q} \rightarrow
 e e^*$ is
\begin{equation}
L = \frac{4\pi}{\Lambda^2}\bar q_L \gamma^\mu q_L \bar E _L \gamma_\mu e_L + h.c. \quad , \nonumber
\end{equation} 
where $E$ denotes the $e^*$ field and $\Lambda$ is the compositeness
 scale. The gauge-mediated (GM) model Lagrangian
 describing the $e^*$ coupling to SM gauge fields is~\cite{baurestar}
\begin{equation}
L = \frac{1}{2\Lambda} \bar E_R \sigma^{\mu \nu} \left[ fg \frac{\vec \tau}{2} \cdot {\vec W}_{\mu \nu} 
+ f'g' \frac{Y}{2} B_{\mu \nu} \right] e_L + h.c. , \nonumber
\end{equation} 
 leading to the reaction  $q \bar q \rightarrow Z / \gamma \rightarrow
 e e^*$. ${\vec W}_{\mu \nu}$ and $B_{\mu \nu}$ are the 
  $SU(2)_L$ and $ U(1)_Y$ field-strength tensors, 
 $g$ and $g'$ are the corresponding electroweak
 couplings, and 
 $f$ and $f'$ are phenomenological parameters where we set $f=f'$.

 Direct searches for $e^*$ production have been 
 performed at HERA by the ZEUS~\cite{zeus}
 and H1~\cite{h1}
 experiments and
 by the LEP2~\cite{lep2,opal} 
 experiments. Mass limits have been set using the GM model only.
 The most stringent LEP limits are set by the OPAL experiment, which 
  has excluded $M_{e^*} < 207$~\gevcc\ 
 for $ f/\Lambda > 10^{-4} $~GeV$^{-1}$ and $M_{e^*} < 103.2$ \gevcc\ for any
 value of
 $f/\Lambda$~\cite{opal}, all at 95\% C.L..
  The most stringent limits from HERA are set by
 the H1 experiment, excluding $M_{e^*} < 280$ \gevcc\ at 95\% C.L. 
 for $f / \Lambda \sim
 0.1$ GeV$^{-1}$~\cite{h1}. In this Letter, we extend the sensitivity to
 higher values of $M_{e^*}$, for $f / \Lambda > 
 0.005$ GeV$^{-1}$. We present the first $e^*$ search in the
 context of the CI
  model, and the first $e^*$ search at a hadron collider. 

 We use 202 pb$^{-1}$ of data collected by the
  CDF II detector~\cite{tdr} during
 2001-2003, from \ppbar\ collisions at $\sqrt s = 1.96$ TeV at the
 Fermilab Tevatron. The detector consists of a magnetic spectrometer with
 silicon and drift chamber trackers,
 surrounded by a time-of-flight system, 
 pre-shower detectors, electromagnetic (EM) and hadronic (Had) calorimeters, 
 and muon detectors. The main components used in this analysis are the 
 central drift chamber
  (COT)~\cite{cot}, the central pre-shower detector~\cite{showermax}
  (for detecting
 photon conversions), and the 
 central~\cite{run1detector} and forward~\cite{plugcal} calorimeters.
 Wire and strip chambers~\cite{showermax} are 
 embedded 
 in the  central EM
 calorimeter to measure transverse shower profiles for $e / \gamma$ 
 identification. 
 The COT, central calorimeter and pre-shower detectors cover the region
 $| \eta | < 1.1$ and the forward calorimeters extend $e / \gamma$ coverage
 to $| \eta | < 2.8$, where $\eta$ is the pseudorapidity.

 We trigger on
 central electron candidates based on high transverse-energy~\cite{perpendicular} EM 
 clusters with associated high transverse-momentum~\cite{perpendicular}
 tracks, with
 an  efficiency (governed by the
  track trigger requirement) of 
  $(96.2 \pm 0.1 )$\%.
 We also use a second electron 
 trigger, with a higher \et\ threshold, but with less
 restrictive identification requirements, which ensures
 $\approx 100$\% efficiency for \et\ $ > 100$ GeV. 
In the offline analysis, 
 we require two fiducial electron candidates (without charge criteria)
 and a photon candidate, each with \et\ $ > 25$ GeV.  We 
 require the isolation $I_{0.4} < 0.1$, where $I_{0.4}$ is the ratio
 of the total 
 calorimeter \et\ around the EM cluster within a radius of $R \equiv
 \sqrt { (\Delta \eta) ^ 2 + (\Delta \phi) ^ 2} = 0.4$ to the cluster
 \et, and $\phi$ is the azimuthal angle.  Longitudinal 
 and lateral shower profiles 
 are required to be consistent with the expectation for EM showers taken 
 from test-beam data. 

 Central electrons are identified by requiring 
 a matching COT track, while 
 central photons are vetoed by 
 a matching COT track with $p_T > ( 1 + 0.005 \times $\et/GeV) \gevc.
 Forward electrons and
 photons are not distinguished from each other
 by using tracking information (in order
 to maximize selection efficiency), but are  
 collectively identified as forward EM objects.
 Events with any dielectron invariant mass in the range 
 $81 < m_{ee} < 101$~\gevcc\ are rejected to suppress $Z ( \rightarrow
 ee ) \gamma$ background. 

We use a  GEANT\cite{geant}-based detector simulation to obtain 
 the offline identification efficiencies. The simulation is validated
  using an unbiased ``probe''
 electron 
 from $Z \rightarrow ee$ events that are triggered and identified
 using the other electron. 
 We measure the central electron  efficiency of 
 $(94.0 \pm 0.3_{\rm stat} )$\% from the data,  compared to 
 $(92.7 \pm 0.1_{\rm stat} )$\%
 from the {\sc pythia}~\cite{pythia} simulation. 
 The simulation of photons is validated by 
 using the EM shower of the probe electron to emulate a photon.
  The measured ``emulated photon''
 efficiency from data 
 (simulation) is $75.5 \% \pm 0.7_{\rm stat} \% $ 
 ($78.3 \% \pm 0.2_{\rm stat} \% $). The simulated efficiency of prompt
 photons is 76\%, showing that the emulated photon is a good model for
 a real photon. 
  The forward EM object efficiency is 
 $89.0 \% \pm 0.6_{\rm stat} \% $ $(90.0 \% \pm 0.6_{\rm stat} \% )$
 in the data (simulation). 
 The inefficiency (due to extraneous energy near the forward EM
 object) decreases 
 with increasing \et, falling below 1\% for 
 \et\ $ > 100$ GeV. 
 Based on the data-simulation comparisons 
 we assign a systematic uncertainty of 1\% (3\%) to the simulated
 central electron (photon) efficiency. 

 We calibrate the EM energy response by
 requiring the measured $Z ( \rightarrow ee ) $ boson mass to agree with the 
  world average~\cite{zmass}. The simulated resolution is tuned using
  the observed
 width of the mass peak. 
  We calculate the full acceptance (including trigger, geometric, kinematic and
 identification efficiencies) 
 using the detector simulation.
 We generate $e e^* \rightarrow e e \gamma$ events using
 {\small PYTHIA}~\cite{pythia} for the CI model, and the 
  {\small LANHEP}~\cite{lanhep}
 and {\small COMPHEP}~\cite{comphep} programs for the GM model.
 The acceptance increases from 15\% at $M_{e^*} = 100$ \gevcc\
 to an asymptotic value of 33\% at high mass, with the largest
 difference between the models of $\approx$5\% at $M_{e^*} = 
 200$ \gevcc. The dominant systematic uncertainties come from 
 identification efficiency (2.6\%), 
 passive material (1.4\%), and 
 parton 
 distribution functions (PDFs) 
 (1.0\%), for a total 
 of 3.7\%. 

\begin{figure}[!htbp]
\begin{center}
\epsfysize = 4.4cm
\epsffile{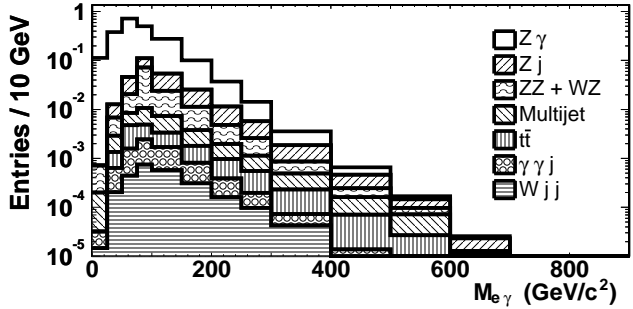}
\vspace*{1mm}
\caption{The cumulative $e\gamma$ mass distribution for all backgrounds. Integrating
 over all masses, the total
 expected number of $e\gamma$ entries is $6.5 \pm 0.1 \; {\rm (stat)} 
 ^{+0.9}_{-0.7} \; {\rm (syst)}$.}
\label{backgrounds}
\end{center}
\end{figure}

 Sources of background, in order of decreasing contribution, are production of
 (1) $Z \gamma \rightarrow ee \gamma$, (2) $Z ( \rightarrow ee) +$~jet where
 the jet is mis-identified as a photon, (3) $WZ \rightarrow eee \nu$ and
  $ZZ \rightarrow eeee$ where
 an electron is mis-identified as a photon, (4) multi-jet events where jets
 are mis-identified as electrons and photons,  (5) $t ( \rightarrow e \nu b ) 
 \bar{t} ( \rightarrow e \nu \bar{b} ) $ with
 energetic photon radiation off the $b$ quarks, (6) 
 $\gamma \gamma +$jet events, and (7)  
 $W (\rightarrow e \nu) + 2$~jets where the jets are mis-identified as an
 electron and a photon.
       
 We estimate the $Z\gamma$, $WZ$, $ZZ$, $t \bar{t}$
  and $\gamma \gamma +$jet backgrounds using
 simulated events, with the {\sc zgamma}~\cite{zgamma} generator for the
 $Z\gamma$ process and {\sc pythia} for the others.  
  Their uncertainties are due to
 integrated luminosity (6\%)~\cite{luminosity}, PDFs (5\%), 
 higher-order QCD corrections (5\%)~\cite{vanneerven}, 
 identification efficiencies (1\%-3\%), passive material (4\%)
 and energy scale and resolution (1\%). 

\begin{table}
\caption{\label{runningSum} Comparison of data and
 integrated background predictions
  above a given cut on the invariant mass of 
 all $e\gamma$ combinations (left) and on the $ee\gamma$ 
 invariant mass (right).}
\begin{tabular}{ccc|ccc}
\multicolumn{3}{c|}{$e\gamma$ combinations}&\multicolumn{3}{c}{events}\\
$m_{e\gamma}$ cut & data & bkg. & $m_{ee\gamma}$ cut & data & bkg. \\
\hline
$>0$ \gevcc\  & 7 & $6.5^{+0.9}_{-0.7} $    & $>0$ \gevcc\ & 3 & $3.0^{+0.4}_{-0.3} $\\
$>50$ \gevcc\ & 7 & $5.3^{+0.8}_{-0.6} $    & $>100$ \gevcc\ & 3 & $2.3^{+0.4}_{-0.3} $\\
$>100$ \gevcc\ & 3 & $2.3^{+0.4}_{-0.3} $    & $>150$ \gevcc\ & 3 & $1.7 \pm 0.3 $\\
$>150$ \gevcc\ & 3 & $0.8^{+0.2}_{-0.1} $    & $>200$ \gevcc\ & 2 & $0.9 \pm 0.2 $ \\
$>200$ \gevcc\ & 2 & $0.31^{+0.10}_{-0.05} $ & $>250$ \gevcc\ & 2 & $0.4 \pm 0.1 $ \\
$>250$ \gevcc\ & 1 & $0.12^{+0.04}_{-0.02} $ & $>300$ \gevcc\ & 2 & $0.18^{+0.06}_{-0.04} $ \\
$>300$ \gevcc\ & 0 & $0.04^{+0.02}_{-0.01} $ & $>350$ \gevcc\ & 0 & $0.08^{+0.03}_{-0.02} $ \\
\end{tabular}
\label{tab:IntegratedBg}
\end{table}

 Backgrounds from $Z +$jet, $W +2$~jet and 
 multi-jet sources are estimated using
 data samples of such events, weighted by the measured ``fake'' rates for jets
 to be misidentified as electrons and photons. The photon fake rate is
 corrected for the prompt photon fraction in the jet sample,
 which is estimated
  using conversion signals observed in
 the calorimeter pre-shower detector. 
 The central electron and photon
 fake rates are $\cal{O}$$(5\times10^{-4})$. 
 The systematic uncertainty in the central photon 
 fake rate ranges from $\sim$50\% at 
 low \et\ (due to variation with $\eta$) to a factor of $\sim$2 at
 high \et\ (due to statistical uncertainty on 
 the prompt photon fraction). 
The fake rate
 for forward
 EM objects is an increasing 
 function of $\eta$ and \et\ with value of $\cal{O}$$(10^{-2})$
 and with systematic uncertainty of a factor of $\sim$2 (due to variation
 with jet sample). All fake rates are
 applied
 as functions of \et, and the forward EM object fake rate is also applied
 as a function of $\eta$. In the $Z-$veto region 
 (~$81 < m_{ee} < 101$ \gevcc)  we
 observe 8 events and predict
  $5.8 \pm 0.1 \; ({\rm stat}) \; ^{+0.9}_{-0.5}
 \; ({\rm syst})$. 
 
For the $e^*$ resonance search, we compare the data with the expected
 background in a sliding window of $\pm 3\sigma$ width on the $e\gamma$ 
 invariant mass distribution, where $\sigma $
  is the RMS of the  $e^*$ mass peak estimated from the simulation.
  All $e\gamma$
 combinations are considered. The RMS is dominated by the detector resolution
 ($\approx$3.5\%)
 over almost the entire $e^*$ parameter space. 
 Figure~\ref{backgrounds} shows the  background predictions 
 for $e\gamma$ combinations. 

\begin{table}
\caption{\label{kinematics} Kinematics of the candidate events. $e, \gamma$, 
 $e'$ and $j$ 
 represent electron, photon, EM cluster and jet 
 respectively. For forward EM objects,
 $e$ and $\gamma$ serve as distinguishing labels only.
  The fractional energy
 resolution for the central and forward calorimeters is given
 by sampling terms of $0.135 \sqrt{{\rm GeV} / E_{\rm T}}$ and 
 $0.16 \sqrt { {\rm GeV} / E}$ respectively, with constant terms of
 $\cal O$(2\%). The $\eta$, $\phi$ and mass resolutions are $\approx$0.005, 
$\approx$0.003 and  $\approx$3.5\% respectively. The jet
 in Event 3 is reconstructed with a cone radius $R = 0.4$, has its energy
 corrected for detector effects,  and has
 energy and $\eta-\phi$ resolutions of $\approx$20\% and $\approx$0.01 
 respectively. }
\begin{tabular}{c|c|c|c}
kinematic & Event 1 & Event 2 & Event 3 \\
\hline
\et$ (e_1)$, charge($e_1$) & 37 GeV, $+$ & 44 GeV, $-$ & 164 GeV, $+$ \\
\et$ (e_2)$, charge($e_2$) & 71  GeV, n.a. & 42  GeV, $-$ & 94  GeV, $-$ \\
\et$ (\gamma)$ & 48  GeV &  46  GeV & 72  GeV \\
$\eta (e_1) , \phi (e_1)$ & $-$1.01, 0.62 & 0.83, 3.64 & $-$0.03, 1.73 \\
$\eta (e_2) , \phi (e_2)$ & 1.27, 4.05 & $-$0.17, 1.96 & 0.46, 5.00 \\
$\eta (\gamma) , \phi (\gamma)$ & $-$1.64, 2.02 & 1.47, 0.92 & $-$0.29, 5.02  \\
$m (e_1e_2)$ & 176  \gevcc\ & 78  \gevcc\ &  256  \gevcc\ \\
$m (e_1\gamma)$ &61  \gevcc\  & 92  \gevcc\  & 219  \gevcc\ \\
$m (e_2\gamma)$ &257  \gevcc\  & 92  \gevcc\ &  64 \gevcc\ \\
$m (e_1e_2\gamma)$ & 318  \gevcc\   &152  \gevcc\ & 343  \gevcc\ \\
\et$ (e'/j)$ & & 26  GeV & 32 GeV \\
$\eta (e'/j) ,  \phi (e'/j)$ & & 1.53, 5.08 & $-$0.50, 3.16\\
$m (e_2e')$ & & 92  \gevcc\ &  
\end{tabular}
\end{table}

We find three candidate events, consistent with our
 predicted background of $3.0 \pm 0.1 \; {\rm (stat)}
 ^{+0.4} _{-0.3} \;
{\rm (syst)} $. The systematic uncertainty receives  equal 
 contributions from the
 uncertainty on the SM backgrounds and the uncertainty on the
 mis-identification backgrounds
 due to the fake rates. 
Comparisons of data and backgrounds are shown in Table~\ref{tab:IntegratedBg}.
 The kinematics of the candidates
  are presented
 in Table~\ref{kinematics}. In Event 1 the forward ``$\gamma$'' 
 has an associated track in the
 silicon detector and is consistent with being a negative electron.
 Event 2 has an additional EM cluster ($e'$) that passes 
 forward selection cuts but marginally fails the isolation cut 
 $(I_{0.4} = 0.107)$. Both forward objects have associated tracks in the
 silicon detector and are consistent with being positive electrons.
 The masses of 
 the $(e_1, \gamma)$
 and $(e_2,e')$ pairs are consistent with
 the event being a $Z(\rightarrow ee)Z(\rightarrow ee)$ candidate.
 
\begin{figure}[!htbp]
%\begin{figure}[b]
\begin{center}
\epsfysize = 4.4cm
\epsffile{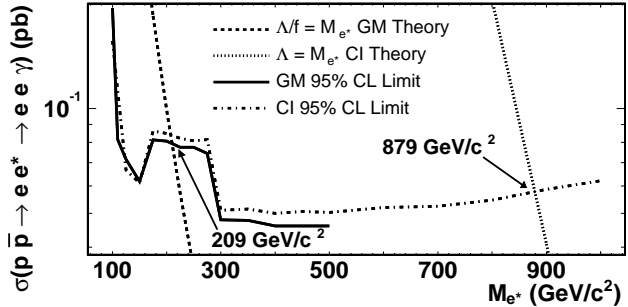}
\vspace*{1mm}
\caption{The experimental cross section $\times$ branching ratio 
limits for the CI and GM models 
 from this analysis,
 compared to the CI model 
 prediction for $ \Lambda = M_{e^*}$
  and the GM model prediction for $ \Lambda / f
 = M_{e^*}$.
  The mass limits are indicated. }
\label{crossSectionLimit}
\end{center}
\end{figure}

\begin{figure}[!htbp]
\begin{center}
\epsfysize = 5.cm
\epsffile{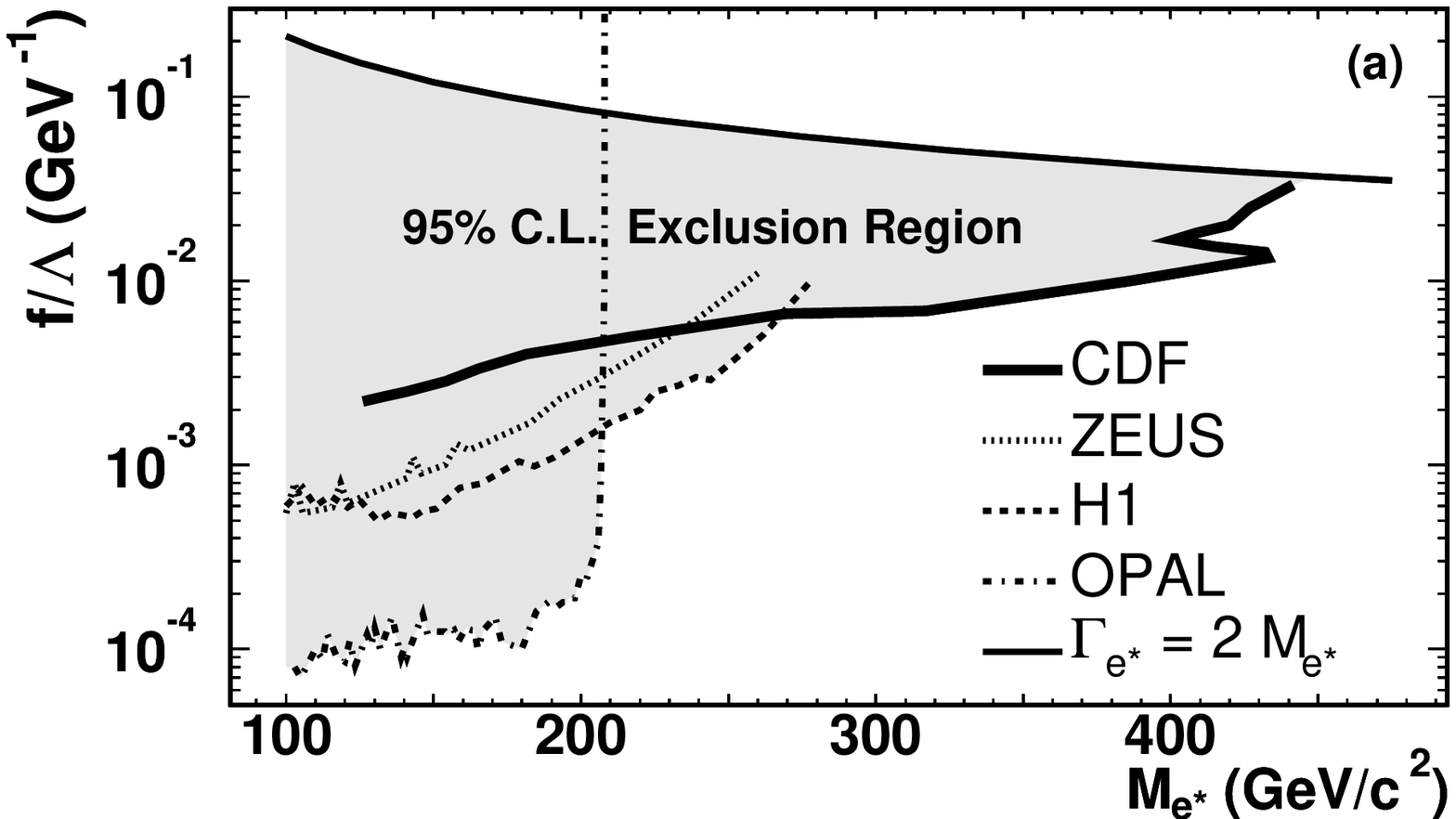}
\epsfysize = 3.5cm
\epsffile{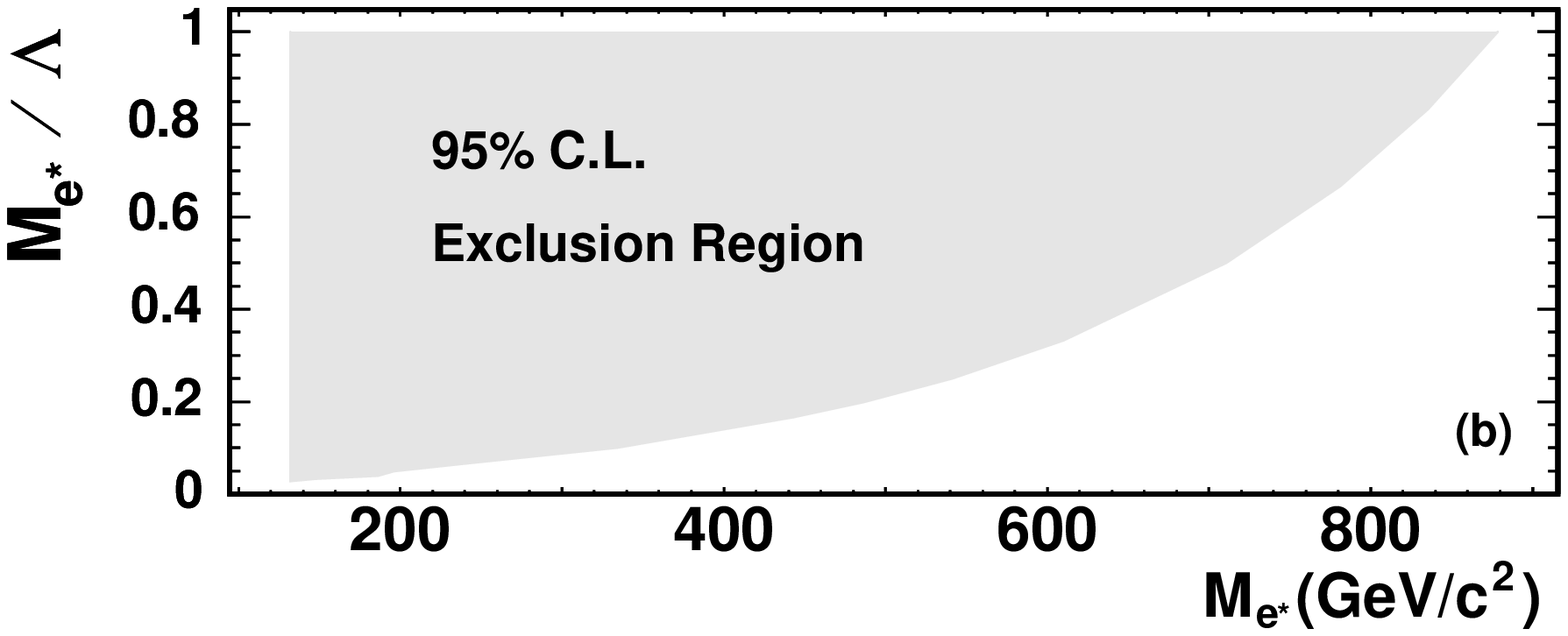}
\vspace*{1mm}
\caption{The 2-D parameter space regions  excluded by this analysis for
 (a) the GM model, along with the current world limits, and (b) the CI model.}
\label{limits}
\end{center}
\end{figure}

 We set limits on $e^*$ production using a Bayesian~\cite{zmass,bayes}
  approach, with a flat prior for the signal and Gaussian priors for the
 acceptance and background uncertainties. 
 The 95\% C.L. upper limits on the cross section $\times$ branching ratio
 (see Fig.~\ref{crossSectionLimit})
 are converted into $e^*$ mass limits by comparison with
 theory~\cite{vanneerven}. For both production models, the  
 $e^*$ decay is prescribed by the GM Lagrangian, which predicts BR$(e^* \rightarrow e 
 \gamma) \approx 0.3$ for $M_{e^*} > 200$~GeV. 
 We include mass-dependent
 uncertainties in the theoretical cross 
 sections due to PDFs (5\%-18\%) and higher-order QCD corrections (7\%-13\%).
  Figure~\ref{limits}
 shows the limits in the parameter space of $f / \Lambda$ ($M_{e^*} / \Lambda$)
 versus $M_{e^*}$ for the GM (CI) model.
The region above the curve labeled 
``$\Gamma_{e^*} = 2 M_{e^*}$'' is unphysical for the GM model, because the
 total  width $\Gamma_{e^*}$ becomes larger than the mass. 

In conclusion, we have presented the results of the first search for
 excited and exotic electrons at a hadron collider. We find three 
 events, consistent with our predicted background.
 In the GM model, we exclude 126~\gevcc\ $< M_{e^*} < 430$ \gevcc\ for
 $f / \Lambda \approx 0.01$ GeV$^{-1}$ at the 95\% C.L., well beyond previous
 limits~\cite{zeus,h1,lep2,opal}.
 We have also presented the first $e^*$ limits in the
 CI model as a function of $M_{e^*}$ and $\Lambda$, excluding
 132~\gevcc\ $ < M_{e^*} < 879$ \gevcc\ for $ \Lambda = M_{e^*}$. 

We are grateful to Alejandro Daleo for providing NNLO cross section
 calculations. 
We thank the Fermilab staff and the technical staffs of the participating institutions for their vital 
contributions. This work was supported by the U.S. Department of Energy and National Science Foundation; 
the Italian Istituto Nazionale di Fisica Nucleare; the Ministry of Education, Culture, Sports, Science 
and Technology of Japan; the Natural Sciences and Engineering Research Council of Canada; the National 
Science Council of the Republic of China; the Swiss National Science Foundation; the A.P. Sloan
 Foundation; the Bundesministerium f\"ur Bildung und Forschung, Germany; the Korean Science and
 Engineering Foundation and the Korean Research Foundation; the Particle Physics and Astronomy Research 
Council and the Royal Society, UK; the Russian Foundation for Basic Research; the Comision 
Interministerial de Ciencia y Tecnologia, Spain; and in part by the European Community's Human 
 Potential Programme under contract HPRN-CT-2002-00292, Probe for New Physics.


\begin{references}
\bibitem{perpendicular}  
``Transverse'' energy ($E_{\rm T}$) and momentum $(p_{\rm T})$ 
 imply the respective components 
 perpendicular to the beam axis. Track $p_{\rm T}$ 
 is obtained from its curvature, and $E_{\rm T}
 = E \; {\rm sin}
 \; \theta $, where $E$ is  the EM cluster energy.
\end{references}
\end{document}